# Surface-Dominated Anomalous Hall Effect in Altermagnetic MnTe Films

Ling-Jie Zhou[1,8], Zi-Jie Yan[1,8], Hongtao Rong[1,8], Yufei Zhao[1,2,8], Pu Xiao[1], Lok-Kan Lai[1], Zhiyuan Xi[1], Ke Wang[3], Tibendra Adhikari[4], Ganesh P. Tiwari[4], Zhong Lin[4,5], Pascal Manuel[6], Fabio Orlandi[6], Dmitry Khalyavin[6], Alexander J. Grutter[7], Chao-Xing Liu[1], Binghai Yan[1,2], and Cui-Zu Chang[1]

[1] Department of Physics, The Pennsylvania State University, University Park, PA 16802, USA

[2] Department of Condensed Matter Physics, Weizmann Institute of Science, Rehovot 7610001, Israel

[3] Materials Research Institute, The Pennsylvania State University, University Park, PA 16802, USA

[4] Department of Physics, Applied Physics and Astronomy, Binghamton University, Binghamton, NY 13902, USA

[5] Materials Science and Engineering Program, Binghamton University, Binghamton, NY 13902, USA

[6] ISIS Neutron and Muon Source, STFC Rutherford Appleton Laboratory, Harwell Campus OX11 0QX, Oxfordshire, UK

[7] Center for Neutron Research, National Institute of Standards and Technology, Gaithersburg, MD 20899, USA

[8] These authors contributed equally: Ling-Jie Zhou, Zi-Jie Yan, Hongtao Rong, and Yufei Zhao

Corresponding authors: binghai.yan@psu.edu (B. Y.); cxc955@psu.edu (C.-Z. C.)

**Abstract: Altermagnets have recently emerged as a new class of magnetic materials that combine compensated magnetic order with spin-split electronic band structures. In this work, we employ molecular beam epitaxy to grow MnTe thin films with controlled thickness on InP(111)A substrates. Using angle-resolved photoemission spectroscopy, we observe a**

**large spin splitting of ≈230 meV for bulk bands well below the Fermi level and identify surface states that cross the Fermi level. Electrical transport measurements show that a robust anomalous Hall (AH) effect persists down to 2 K and an AH sign reversal occurs near 175 K. By systematically tuning film thickness, growth conditions, and interfacial structure, we demonstrate that the AH response in MnTe films originates from surface states rather than bulk bands. Our first-principles calculations reveal that this surface-dominated AH effect is imprinted by the bulk altermagnetic order and remains unchanged for terminations with opposite Mn magnetic orientations. Our results establish a unique surface transport probe of bulk altermagnetism, demonstrate interface engineering as an effective route to generate and control the AH effect in altermagnets, and provide a unified understanding of the AH response in altermagnetic MnTe films.**

**Main text:** Magnetic materials have traditionally been classified into two categories: ferromagnets and antiferromagnets. Ferromagnets possess a net macroscopic magnetization that breaks time-reversal symmetry and produces spin-polarized electronic bands, enabling phenomena such as the anomalous Hall (AH) effect. In contrast, antiferromagnets host compensated magnetic moments, resulting in vanishing net magnetization and, in most cases, spin-degenerate electronic structures in momentum space. While ferromagnets have long dominated spintronics because of their ease of electrical detection and control, antiferromagnets offer advantages such as a vanishing stray field, robustness against external magnetic fields, and ultrafast spin dynamics. Bridging these two paradigms has been a long-standing goal in condensed-matter physics. Recently, this dichotomy has been fundamentally expanded by the identification of altermagnetism, a distinct magnetic phase in which compensated collinear magnetic order coexists with a large momentum-dependent spin splitting [1-8]. In altermagnets, opposite spin sublattices are not related by inversion or translation symmetries. Instead, they are connected by other crystal symmetry operations, such as rotations, mirror reflections, glide planes, and screw axes, resulting in vanishing net magnetization but spin-polarized band structures and finite Berry curvature, even when spin-orbit coupling (SOC) is negligible [1-8]. These unique properties position altermagnets as a new platform for exploring magnetotransport phenomena beyond conventional ferromagnets and antiferromagnets [1-15].

Altermagnetic band splitting has been directly observed using angle-resolved photoemission spectroscopy (ARPES) measurements on several materials, including $\alpha$-MnTe, CrSb, Rb-deintercalated $V_2Te_2O$, and $KV_2Se_2O$ [16-26], providing unambiguous momentum-resolved evidence for this new magnetic order. Complementary electrical transport measurements have revealed a spontaneous AH effect in $\alpha$-MnTe [27], indicating the presence of nonzero Berry curvature despite compensated magnetization. Unlike metallic CrSb [28-30], $\alpha$-MnTe is a semiconductor with a band gap of $\approx$1.4 eV [27,31-34]. It has emerged as a promising altermagnetic platform for semiconducting spintronic applications owing to its high *Néel* temperature near room temperature [35,36], large spin splitting on the order of $\approx$100 meV [16,17], and excellent compatibility with molecular beam epitaxy (MBE) growth. However, to date, experimental reports on the AH effect in $\alpha$-MnTe have exhibited strong sample dependence [27,37-44], including sign reversals and temperature-restricted behavior, leaving the microscopic origin of the AH response in $\alpha$-MnTe unresolved.

In this work, we employ MBE to grow *d*-unit-cell (UC) thick $\alpha$-MnTe films on lattice-matched InP(111)A substrates. ARPES measurements reveal a large spin splitting of $\approx$230 meV for the bulk valence band near the Fermi level $E_F$, directly confirming the altermagnetic band structure. Importantly, we identify surface states that cross $E_F$ and dominate the low-energy electronic band structure. Electrical transport measurements uncover a robust AH effect persisting down to 2 K and a pronounced sign reversal near 175 K. The longitudinal resistance of α-MnTe films shows a weak thickness dependence for $d \geq 5$, indicating surface-dominated conduction. By systematically varying film thickness, growth conditions, and interfacial structure, we identify two competing AH contributions of opposite signs, which we attribute to distinct surface channels at the top Te/MnTe and bottom MnTe/InP interfaces. Our first-principles calculations identify distinct surface states at these two interfaces with opposite AH conductance and reveal their intimate connection to the bulk altermagnetic order. These findings establish interface engineering as a key mechanism for inducing Berry curvature and generating the AH effect in α-MnTe, providing a unified explanation that reconciles seemingly inconsistent results from prior experiments[27,37-40].

α-MnTe crystallizes in a hexagonal NiAs-type structure with space group $P6_3/mmc$. The adjacent Te atomic layers are related by a mirror operation $M_z$ or a nonsymmorphic screw-axis rotation $C_{6z}t_{1/2}$, where $t_{1/2}$ denotes a half-UC translation along the $c$ axis, rather than by inversion $P$ or translation $t_{1/2}$ alone (Fig. 1a,b). When the spin configuration is taken into account, $M_z$ or $C_{6z}t_{1/2}$ relates opposite spin sublattices, giving rise to a $g$-wave spin splitting even in the absence of SOC [2-4,16]. When the in-plane magnetic moments align with the next-nearest-neighbor Mn-Mn direction, a finite Berry curvature is allowed, enabling the AH effect [27]. As noted above, although the AH effect has been reported in α-MnTe films and single crystals, its microscopic origin remains debated [27,37-44]. Positive AH resistance with a soft ferromagnetic-like hysteresis loop has been observed in MBE-grown α-MnTe films at $T \approx 150$ K [27,37-39], whereas MnTe single crystals exhibit highly sample-dependent AH behavior, with reported AH signs that are positive, negative, or even reversed [41-44]. Moreover, for MBE-grown α-MnTe films, the AH effect is prominent only for 150 K < $T$ < 250 K and strongly suppressed at lower $T$ [27,37-39], raising the possibility that extrinsic factors, such as structural anomalies or excess Mn dopants, rather than intrinsic Berry curvature of the bulk altermagnetic ground state, may play a dominant role [37,38].

We first optimize the MBE growth recipes of α-MnTe films on InP(111)A substrates [45] and then perform systematic structural characterizations. The lattice constant of InP is $a_0$ = 0.590 nm. For the InP(111)A surface, the corresponding in-plane lattice parameter $a_0/\sqrt{2} \approx$ 0.417 nm closely matches that of α-MnTe ($a$ = 0.416 nm). Due to excellent lattice compatibility, sharp, streaky reflection high-energy electron diffraction (RHEED) patterns are observed throughout the MBE growth process (Figs. S1 and S2) [45]. Cross-sectional scanning transmission electron microscopy (STEM) images of the MnTe films with $d$ = 20 and $d$ = 50 reveal an atomically sharp interface between MnTe and InP(111)A (Figs. 1c, S3, and S4). In addition, STEM measurements show that the first few Te capping layers form an atomically sharp interface with the MnTe film (Figs. S3b and S4b) [45]. An atomic force microscopy image of an uncapped $d$ = 40 film reveals a uniform surface morphology with a root-mean-square roughness of ≈0.856 nm (Fig. 1d). Neutron diffraction measurements performed on a $d$ = 230 film show the emergence of the (001) magnetic Bragg peak for $T$ < 300 K, confirming the onset of antiferromagnetic coupling between the two Mn sublattices (Fig. S5). Fitting the $T$ dependence of the peak intensity yields a *Néel* temperature $T_N$ = (303.9 ± 0.5) K (Figs. 1e and S6). X-ray diffraction (XRD) spectra of three MnTe films with $d$ = 50, 140, and 230 confirm the α-MnTe phase (Figs. 1f and S7), and the rocking curve of the MnTe (004) diffraction peak exhibits a full width at half maximum (FWHM) of ≈0.18°, indicating excellent epitaxial growth of α-MnTe on InP(111)A (Fig. 1g). The coexistence of the NiAs-type lattice and the A-type antiferromagnetic structure satisfies the essential symmetry requirements for altermagnetic order, specifically, broken $PT$ symmetry and the absence of anti-translation $t_{1/2}T$ symmetry, thereby establishing our MBE-grown MnTe films as an ideal platform for exploring the link between spin-split band structures and the AH effect [2,3].

Next, we perform ARPES measurements on two MnTe films with $d = 50$ and pristine surfaces using He-I$_\alpha$ photons with $h\nu = 21.218$ eV (Figs. 2, S8, and S9). As established in prior studies [16,17], this photon energy probes a $k_z$ value slightly away from $k_z$=0. Figure 2a shows the 3D bulk Brillouin zone and its surface projection. Figure 2b shows the second-derivative Fermi surface and constant-energy contours of the MnTe film at binding energies $E_B = 0.1$ eV and $E_B = 0.2$ eV. At $E_B = 0$, the Fermi surface consists of two circular hole-like pockets centered at the Γ point (labeled α and $\alpha_1$). As $E_B$ increases, the α and $\alpha_1$ pockets expand. In addition, six elliptical pockets surrounding the Γ point (labeled $\beta_1$) emerge, with their long axes aligned along the Γ-K direction. The band dispersion measured along the K-Γ-K direction reveals two bands ($\beta_1$ and $\beta_2$) crossing at the Γ point near $E_F$ (Fig. 2c). Weak surface states α and $\alpha_1$ cross $E_F$ and are more clearly resolved in the second-derivative band map (Fig. 2g). Besides $\beta_1$, $\beta_2$, α, and $\alpha_1$ bands, the ARPES band dispersion shows several broad hole-like bands at the Γ point, located at ≈0.5 eV (labeled γ), ≈ 1 eV (δ), and ≈1.2 eV (ε) below $E_F$ (Fig. 2c). The α and $\alpha_1$ bands originate from surface states. In contrast, all other bands are derived from bulk states. Both the bulk band structure and the surface states observed experimentally are well reproduced by our first-principles calculations on MnTe with a top Te termination (Fig. 2d,f,h). We next focus on the spin-split bands near $E_F$. The second-derivative band map shows two well-separated $\beta_1$ and $\beta_2$ bands, with a maximum energy splitting $\Delta E$ of ≈230 meV (Fig. 2e), much larger than the values reported in prior studies [16,17]. This large splitting, arising from the interplay between altermagnetism and SOC, is well reproduced by our first-principles calculations (Fig. 2f) and directly reflects altermagnetism [16,17]. The $\alpha$ and $\alpha_1$ surface states are clearly resolved in the second-derivative band map with respect to the momentum of the band dispersion (Fig. 2g) and agree well with the calculated band structure (Fig. 2h). Moreover, similar band dispersions are observed at a photon energy of $h\nu = 11$ eV [38] but are absent in soft X-ray ARPES measurements [16], further supporting the surface-state origin of the $\alpha$ and $\alpha_1$ bands. Therefore, our ARPES results provide compelling momentum-resolved evidence for bulk altermagnetism in MBE-grown $\alpha$-MnTe films on InP(111)A. Furthermore, these results indicate that transport is dominated by surface states rather than bulk bands, as the top bulk valence band lies $\approx$80 meV below $E_F$.

To probe the connection between bulk altermagnetism and the AH effect, we perform electrical transport measurements on $\alpha$-MnTe films on InP(111)A with $1 \leq d \leq 230$ (Figs. 3, 4, and S10 to S18). Figure 3a shows the $T$ dependence of the sheet longitudinal resistance $\rho_{xx}$. For $d = 1$ and $d = 2$, $\rho_{xx}$ exhibits insulating behavior and diverges at low $T$. In contrast, the $d = 5$ film shows a substantially reduced $\rho_{xx}$ and a pronounced hump near $T \approx 280$ K, associated with the paramagnetic-to-antiferromagnetic phase transition (i.e., $T_N$). For films with $5 \leq d \leq 230$, $\rho_{xx}$ exhibits only a weak dependence on $d$. In this regime, the films show a metallic $T$ dependence over 50 K < $T$ < 280 K, in stark contrast to the behavior expected for a doped semiconductor. As T decreases toward 50 K, a metal-to-insulator transition is observed. Figure 3b shows the $T$ dependence of the normalized longitudinal resistance $\rho_{xx,norm} = \rho_{xx}/\rho_{xx,\min}$. A Kondo-like transport behavior, manifested as a logarithmic upturn in $\rho_{xx}$ at low $T$ [46-49], likely arises from magnetic impurities, such as excess Mn dopants [38]. We further define $T_{\min}$ as the temperature at which $\rho_{xx}$ reaches its minimum and find that $T_{\min}$ increases monotonically with $d$ (Fig. 3c). More experiments are required to clarify the microscopic origin of the $d$-dependent $T_{\min}$ behavior observed in these MBE-grown α-MnTe films. Figure 3d shows the $d$-dependence of $\rho_{xx}$ measured at $T$ = 2 K [denoted as $\rho_{xx}$ ($T$=2K)]. A pronounced reduction in $\rho_{xx}$ ($T$=2K) is observed as $d$ increases from 1 to 5, followed by a nearly $d$-independent resistance for $d \geq 5$. The weak $d$ dependence of $\rho_{xx}$ ($T$=2K), together with its slight increase at larger $d$, indicates that electrical transport in MBE-grown α-MnTe films on InP(111)A is dominated by surface states rather than bulk states [37]. The surface-dominated conduction is consistent with our ARPES results, which show that only the α and $\alpha_1$ surface states cross $E_F$ (Fig. 2b,c,g).

Next, we investigate the magnetotransport properties of the $d$ = 230 film. For 250 K ≤ $T$ ≤ 300 K, the sample exhibits a linear Hall trace (Fig. S10) [45]. Upon further cooling, a soft hysteresis loop emerges at $T$ = 200 K and persists down to $T$ = 2 K. Figure 4a shows the out-of-plane magnetic-field $\mu_0 H$ dependence of the Hall resistance $\rho_{yx}$ for $T \leq 200$ K, after subtracting the linear contribution from the ordinary Hall effect. At $T$ = 200 K, the zero-magnetic-field Hall resistance $\rho_{yx}(0)$ after a positive-magnetic-field training is ≈2.94 Ω. As $T$ decreases to $T$ = 150 K, $\rho_{yx}(0)$ switches sign and becomes ≈ - 4.47 Ω. The AH sign reversal occurs for 150 K < $T$ < 200 K, and the shape of the AH hysteresis loop for $T \leq 150$ K indicates the coexistence of the two AH contributions with opposite AH signs in the $d$ = 230 film [50-54]. As $T$ further decreases, $\rho_{yx}(0)$ decreases monotonically, reaching - 7.71 Ω at $T$ = 100 K, - 10.45 Ω at $T$ = 40 K, and - 23.44 Ω at $T$ = 2 K. At $T$ = 2 K, the Hall trace evolves into a sharp hysteresis loop with a small residual hump feature near the large coercive field $\mu_0 H_c$ ≈ ± 6.0 T, reflecting the hard-ferromagnet-like nature of the $d$ = 230 film. All these observations indicate that the positive AH component dominates for $T \geq 200$ K. In contrast, the negative AH component dominates at $T$ = 2 K. For 2 K < $T$ < 200 K, the positive and negative AH components compete. Their opposite AH signs and distinct $\mu_0 H_c$ suggest that they may arise from two opposite surfaces. Interface-induced strain may modify the magnetic anisotropy of MnTe, thereby tuning $\mu_0 H_c$ and reshaping the AH hysteresis loop [42,43].

To further clarify the evolution of the AH effect, we examine the $\mu_0 H$ dependence of $\rho_{yx}$ for α-MnTe films with $5 \leq d \leq 230$ at $T = 2$ K (Fig. 4b) and $T = T^*$ (Fig. 4c). Here, we define $T^*$ as the temperature at which the positive $\rho_{yx}(0)$ reaches its maximum (Fig. 4d). At $T = 2$ K, a well-defined hysteresis loop with a negative AH sign is observed for $d \geq 20$. In contrast, it gradually shrinks as $d$ decreases and becomes negligible for $d$ = 10 and $d$ = 5. $\rho_{yx}(0)$ at $T$ = 2 K [denoted as $\rho_{yx}(0)(T$=2K)] is found to be ≈ - 23.4 Ω for $d$ = 230, ≈ - 19.3 Ω for $d$ = 50, ≈ - 22.3 Ω for $d$ = 40, and ≈ - 13.5 Ω for $d$ = 20 (Fig. 4b). In contrast, a pronounced hysteresis loop with a positive AH sign is observed for all samples at $T = T^*$, while it gradually expands as $d$ decreases. $\rho_{yx}(0)$ at $T = T^*$ [denoted as $\rho_{yx}(0)(T=T^*)$] is found to be ≈2.9 Ω for $d$ = 230, ≈5.4 Ω for $d$ = 50, ≈7.4 Ω for $d$ = 40, ≈8.6 Ω for $d$ = 20, ≈22.3 Ω for $d$ = 10, and ≈41.3 Ω for $d$ = 5 (Fig. 4c). As $d$ increases from 5 to 230, the appearance of a hysteresis loop with a negative AH sign at $T$ = 2 K, together with the suppression of a hysteresis loop with a positive AH sign at $T = T^*$, further indicates the presence of two competing AH contributions with opposite AH signs in our MBE-grown α-MnTe films on InP(111)A. Figure 4d shows the $T$ dependence of $\rho_{yx}(0)$ for α-MnTe films with $1 \leq d \leq 230$. For $d$ = 230, 50, and 40, $\rho_{yx}(0)$ becomes finite at $T \approx 225$ K. It reaches a maximum at $T^*$ = 200 K. Upon further cooling, $\rho_{yx}(0)$ reverses sign at $T \approx 175$ K and then decreases monotonically as $T$ decreases. For $d$ = 20, $T^*$ remains at 200 K, while $\rho_{yx}(0)$ reverses sign at $T \approx 50$ K. For $d \leq 10$, the positive AH component is significantly enhanced, whereas the negative AH component is strongly suppressed at low temperatures. $\rho_{yx}(0)$ reaches its maximum at $T^*$ = 260 K for $d$ = 10, $T^*$ = 250 K for $d$ = 5, and $T^*$ = 175 K for $d$ = 2, respectively. In contrast, for $d$ = 1, no AH effect is observed, indicating the absence of long-range magnetic order in a single UC α-MnTe layer (Fig. 4d).

Finally, we verify the surface origin of the AH effect using first-principles calculations. We construct a slab model for a MnTe film on an InP(111)A substrate with a Te capping layer (Fig. 5a). By identifying the surface states on the top and bottom surfaces, we evaluate their contributions to AH conductance $\sigma_{xy}$ (Fig. 5b-d). The bottom surface bonds to the InP substrate via interfacial Te atoms, whereas the top surface is Te-terminated, as confirmed by the consistency between our ARPES measurements and first-principles calculations (Fig. 2g,h). The Te capping layer is modeled by adding an extra Te layer on the top Te-terminated surface [45]. Although both surfaces are intrinsically Te-terminated, the adjacent materials perturb their surface states in markedly different ways. The resulting top- and bottom-surface states are highlighted by dotted lines in Fig. 5c,d. At the bottom interface, the wide-bandgap InP substrate ($E_g \approx$ 1.42 eV) [55] primarily donates a fraction of an electron per interfacial atom, leaving the surface-state dispersion largely intact and close to that observed by ARPES for the pristine surface (Fig. 2g). In contrast, at the top surface, the narrow-bandgap and chemically similar Te capping layer strongly hybridizes with the surface states, leading to substantial band reconstruction. In our first-principles calculations, the charge-neutral point is fixed at $E = 0$ eV, and the experimental $E_F$ determined from ARPES is slightly above the bulk bands, at $E \approx$ -0.2 eV. At $E \approx$ -0.2 eV, the top Te/MnTe and bottom MnTe/InP surface bands contribute - 8.4 μS and + 0.7 μS to $\sigma_{xy}$, respectively (Fig. 5b). The calculated surface-state contributions to $\sigma_{xy}$ are comparable in order of magnitude to the experimentally extracted values of - 2 μS < $\sigma_{xy}$ < 2 μS (Fig. S18a) [45]. Our recent theoretical calculations for a Te-terminated MnTe slab further demonstrate that moderate shifts in $E_F$ do not reverse the sign of $\sigma_{xy}$ [56]. In contrast, even taking the upper bound of the reported bulk $\sigma_{xy}$ of ≈ $10^{-2}$ S/cm [42], this corresponds to ≈ 6.71 × $10^{-4}$ μS per UC in 2D. To reach $\sigma_{xy}$ ≈ 1 μS, a minimum thickness of $d \approx 1500$ would be required, far exceeding the thickness of all MnTe films studied in this work ($1 \leqslant d \leq 230$). Therefore, we demonstrate that the AH response in our MnTe films is dominated by surface contributions, even if the bulk is doped. To probe the sensitivity of the surface states to the local environment, we passivate the top surface with hydrogen atoms to simulate oxidation (Fig. S20). This passivation shifts the top surface states downward into the bulk bands and away from $E_F$, while leaving the bottom surface states essentially unchanged [45]. This behavior is consistent with the experimental observation that positive and negative AH components coexist and are highly sensitive to the interfacial environment (Figs. S22 and S23) [45]. Moreover, our MnTe films are in the strong-disorder regime, characterized by low carrier mobility $\mu < 100$ $cm^2 \cdot V^{-1} \cdot s^{-1}$ (Fig. S18c) [45], where extrinsic AH mechanisms are strongly suppressed. This suggests that the observed AH effect is predominantly intrinsic and originates from the Berry curvature of the electronic bands [57].

The surface-dominated AH effect is an intrinsic manifestation of bulk altermagnetism, even though the bulk bands do not cross $E_F$. Our first-principles calculations show that, for a fixed Te/MnTe or MnTe/InP interface structure, the AH sign is independent of the magnetization direction of the surface MnTe layer and instead governed by the bulk altermagnetic order. To explicitly demonstrate this, we focus on the top Te/MnTe interface and compare three different MnTe termination configurations (i), (ii), and (iii) in Fig. 5e. Configurations (i) and (ii) share the same bulk altermagnetic order (see dashed black boxes in Fig. 5e) but are terminated at different MnTe layers with opposite magnetization directions. Because these two configurations can be related by the two-fold rotation $C_{2z}$, which preserves $\sigma_{xy}$, they exhibit the same $\sigma_{xy}$. In contrast, Configurations (i) and (iii) have opposite bulk altermagnetic orders but are terminated at the MnTe layer with the same magnetization direction. We find that these two configurations exhibit opposite $\sigma_{xy}$, since they are related by the combined $C_{2z}T$ symmetry. This symmetry-based comparison unambiguously demonstrates that the AH sign is determined by the bulk altermagnetic order, rather than by the magnetization direction of the surface MnTe layer, even though the AH effect originates from the top and bottom surface layers.

To summarize, we demonstrate the surface-dominated AH effect in MBE-grown $\alpha$-MnTe films on InP(111)A that host bulk altermagnetism. A spin-split band structure induced by bulk altermagnetism is observed, with an energy splitting of $\approx$230 meV well below $E_F$. We find that only surface states cross $E_F$. Consistent with this observation, electrical transport measurements indicate surface-dominated conduction. By systematically varying film thickness, growth conditions, and interfacial structure, we identify two competing AH contributions with opposite AH signs, originating from distinct surface channels at the top Te/MnTe and bottom MnTe/InP interfaces. Our first-principles calculations confirm the coexistence of these two surface channels and reveal that their AH contributions are governed by the combined effects of surface/interface chemical environment and bulk altermagnetism, rather than the magnetization direction of the surface MnTe layer. These findings establish a surface-sensitive probe of bulk altermagnetism, demonstrate interface engineering as an effective route to inducing the AH effect in MBE-grown MnTe films, and lay the foundation for developing altermagnet-based electronic and spintronic devices.

**Acknowledgments:** We thank C. Du, N. Samarth, W. Wu, D. Xiao, and J. Zhu for helpful discussions. This project is primarily supported by the ONR Award (N000142412133), including MBE growth and PPMS(9T) measurements. ARPES and PPMS(14T) measurements and theoretical calculations are supported by the Penn State MRSEC seed project for Nanoscale

Science (DMR-2011839). Sample characterization is supported by the NSF grant (DMR-2241327) and the ARO award (W911NF2210159). TA, GPT, and ZL acknowledge start-up support from Binghamton University. Partial support for AJG is provided by the Center for High Resolution Neutron Scattering, a partnership between the NIST and the NSF under Agreement No. DMR-2010792. B.Y. acknowledges the support from the Israel Science Foundation (ISF: 2974/23). This work identifies certain commercial equipment, instruments, software, or materials to specify the experimental procedure adequately. Such identifications are not intended to imply recommendation or endorsement by NIST, nor are they intended to imply that the materials or equipment identified are necessarily the best available for the purpose. The authors acknowledge the Science and Technology Facilities Council (STFC), part of UKRI, for providing beam time on the WISH diffractometer under proposal No. RB2510207. CZC acknowledges the support from the Gordon and Betty Moore Foundation's EPiQS Initiative (GBMF9063 to C.-Z. C).

**Data availability**: The data that support the findings of this article are openly available [58].

**Figures and figure captions:**

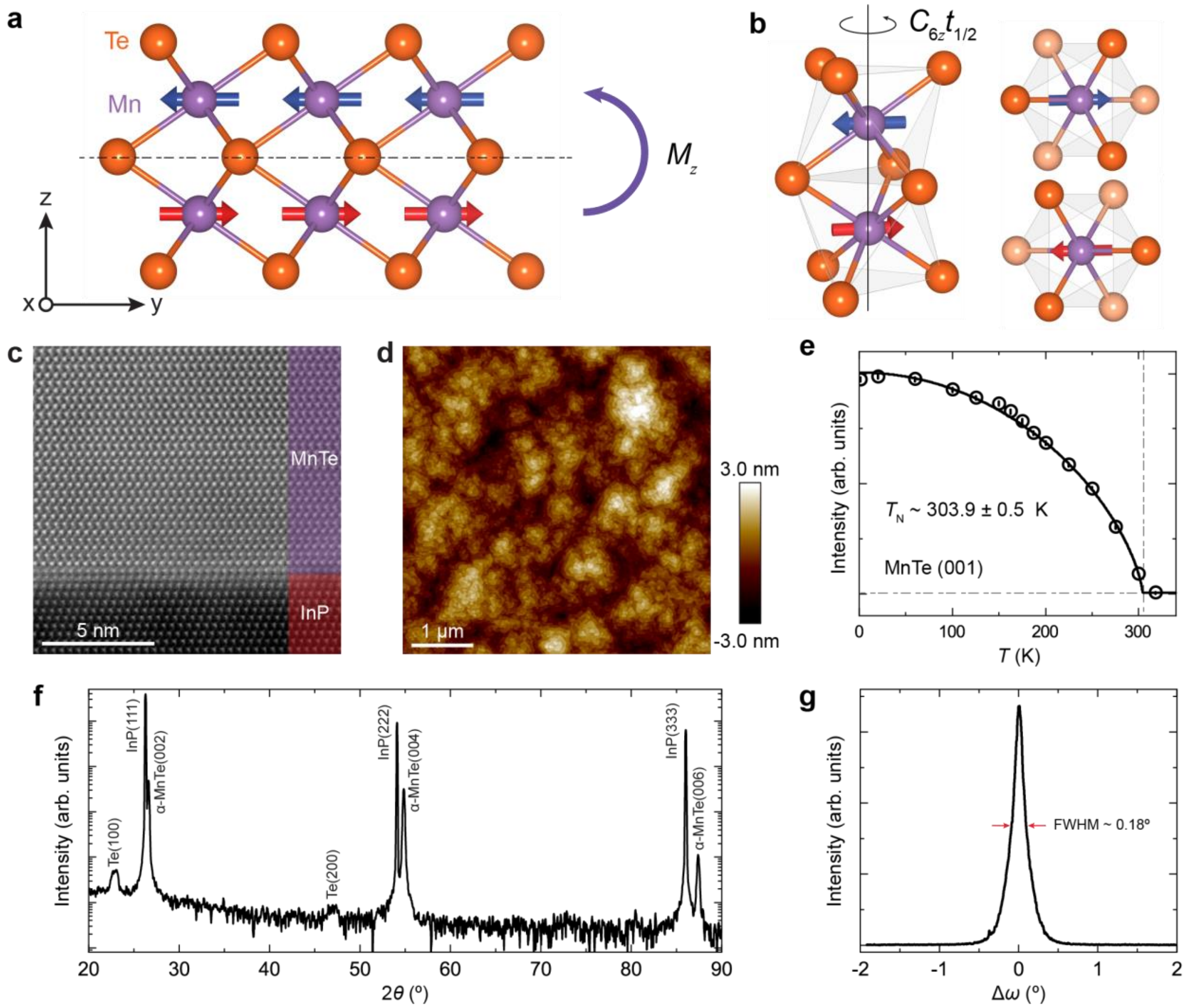

**Fig. 1 | Characterization of epitaxial MnTe films on InP(111)A substrates. a, b**, Crystal and magnetic structure of MnTe. $M_z$ or $C_{6z}t_{1/2}$ relates opposite spin sublattices. $M_z$ is the mirror operation across the basal (001) plane. $C_{6z}$ is a six-fold rotation operation, and $t_{1/2}$ is a half-UC translation along the *c*-axis. **c**, STEM image of a $d = 20$ film. **d**, Atomic force microscope image taken on an uncapped $d = 40$ film. **e,** *T*-dependent (001) peak intensity of a $d = 230$ film measured by neutron diffraction. The fit yields $T_N = (303.9 \pm 0.5)$ K. Error bars represent ±1 standard deviation. **f**, XRD spectrum of a $d = 140$ film. **g**, Corresponding rocking curve of the MnTe (004) peak with a FWHM of ≈0.18°.

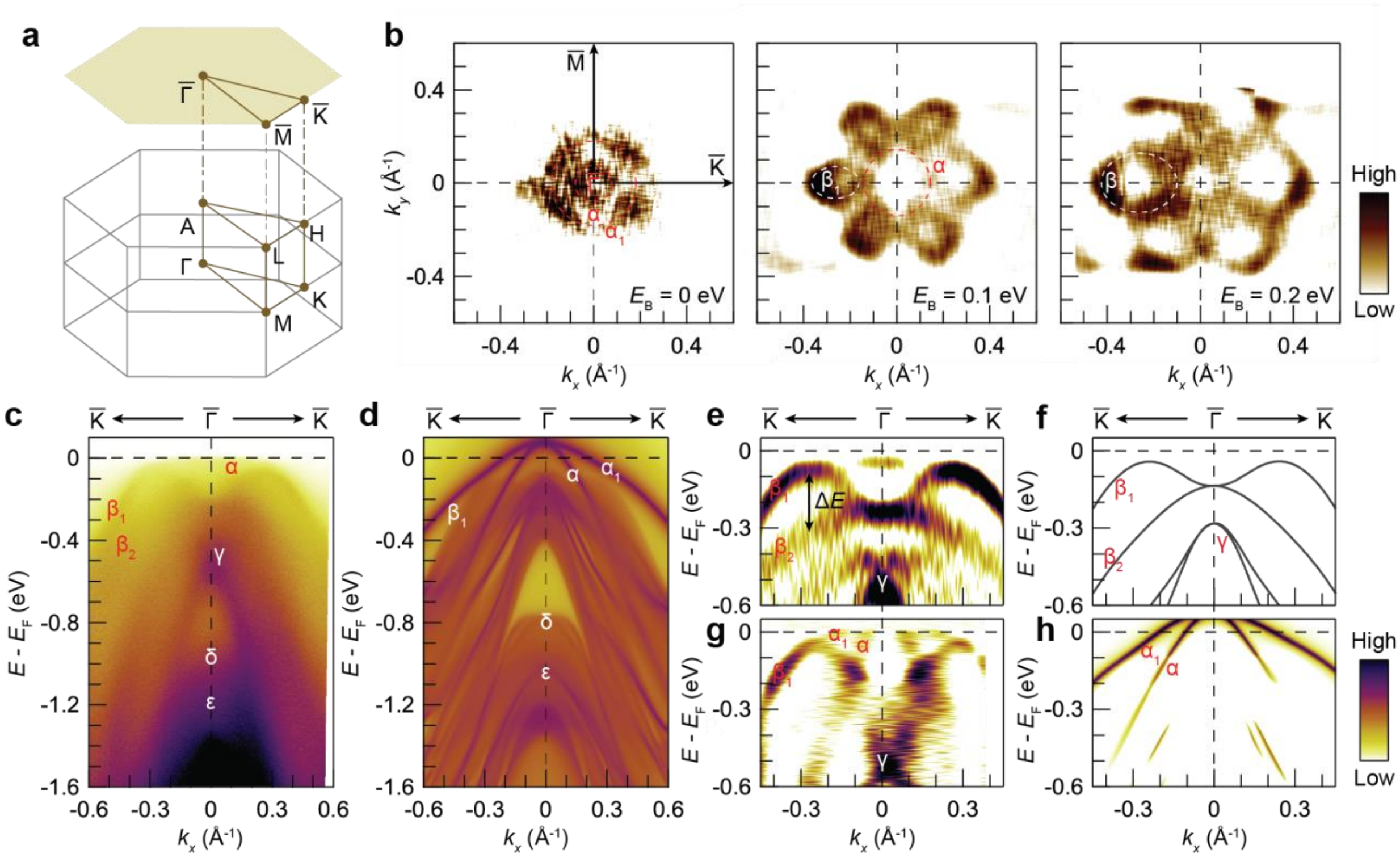

**Fig. 2 | Surface states and spin-split bands in epitaxial MnTe films with $d$ = 50. a**, 3D bulk Brillouin zone of MnTe. The yellow plane at the top represents the surface-projected Brillouin zone. **b**, Second-derivative constant-energy contours at the Fermi surface (left), $E_B$ = 0.1 eV (middle), and $E_B$ = 0.2 eV (right). The second-derivative Fermi surface is obtained by integrating the spectral weight within a [-25, 25] meV window, and the other second-derivative constant-energy contours are obtained by integrating the spectral weight within a [-20, 20] meV window with respect to the corresponding $E_B$. **c**, ARPES band dispersion measured along the K-Γ-K direction. **d**, Calculated surface states and bulk bands of MnTe with a Te-terminated top surface along the K-Γ-K direction in the altermagnetic state with SOC. **e**, The second-derivative band map with respect to the energy of the band structure. The spin-split $\beta_1$ and $\beta_2$ bands are resolved. **f**, Calculated bulk bands along the K-Γ-K direction. **g**, The second-derivative band map with respect to the momentum of the band structure. The $\alpha$ and $\alpha_1$ surface states are resolved. **h**, The same calculated electronic band structure in (**d**), shown with a different color scale to highlight the surface states ($\alpha$ and $\alpha_1$). All ARPES measurements are performed at $T$ = 12 K.

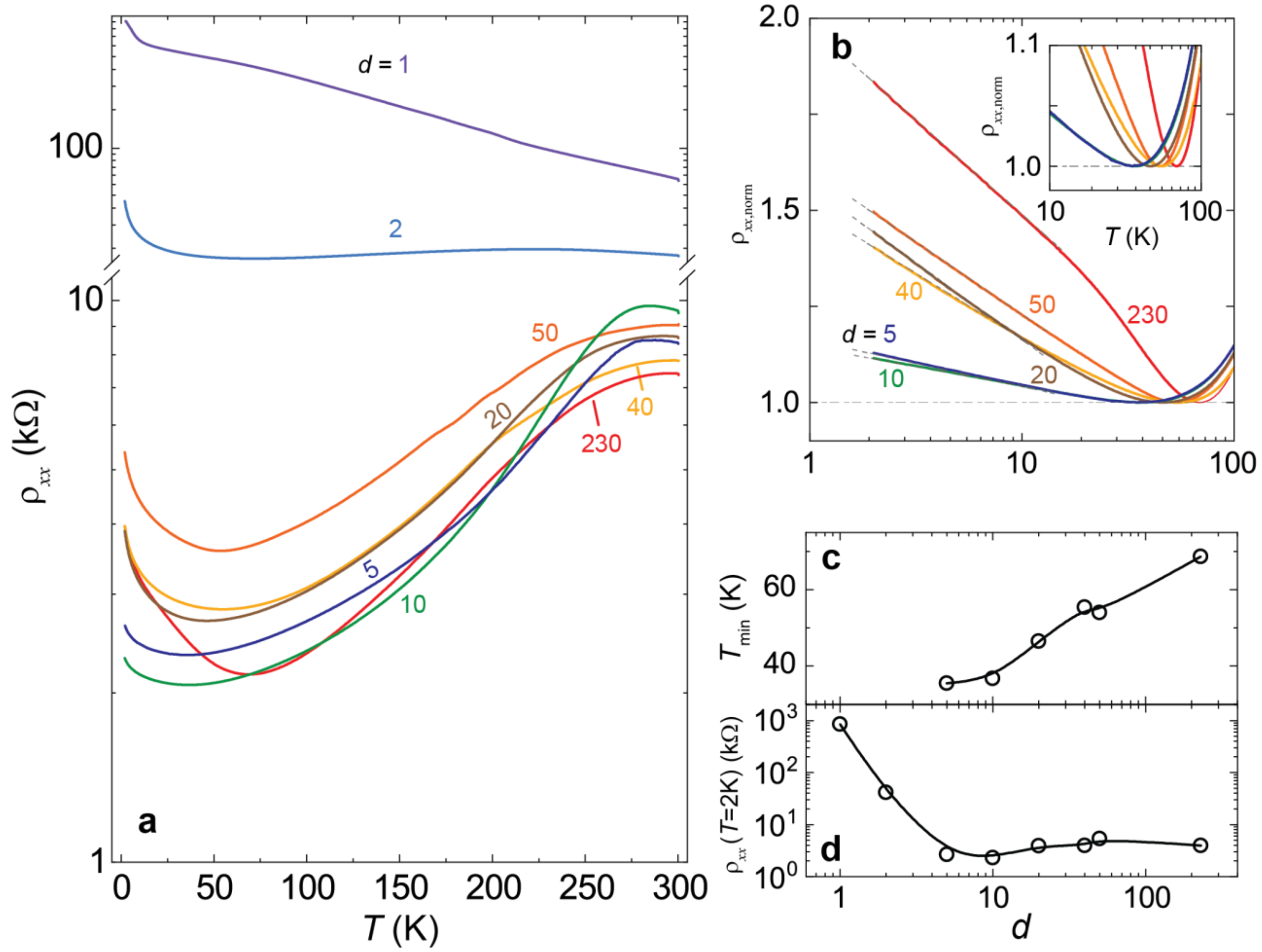

**Fig. 3 | Surface conduction in epitaxial MnTe films**. **a**, $T$-dependent $\rho_{xx}$ of MnTe films with $d$ = 1, 2, 5, 10, 20, 40, 50, and 230. **b**, $T$ dependence of the normalized $\rho_{xx}/\rho_{xx,\min}$ for MnTe films with $d$ = 5, 10, 20, 40, 50, and 230. Inset: Enlarged view near $\rho_{xx,\min}$. The dashed lines indicate the logarithmic increase at low $T$. **c**, $d$ dependence of $T_{\min}$, defined as the temperature at which $\rho_{xx}$ reaches its minimum. **d**, $d$-dependence of $\rho_{xx}$ ($T$=2K).

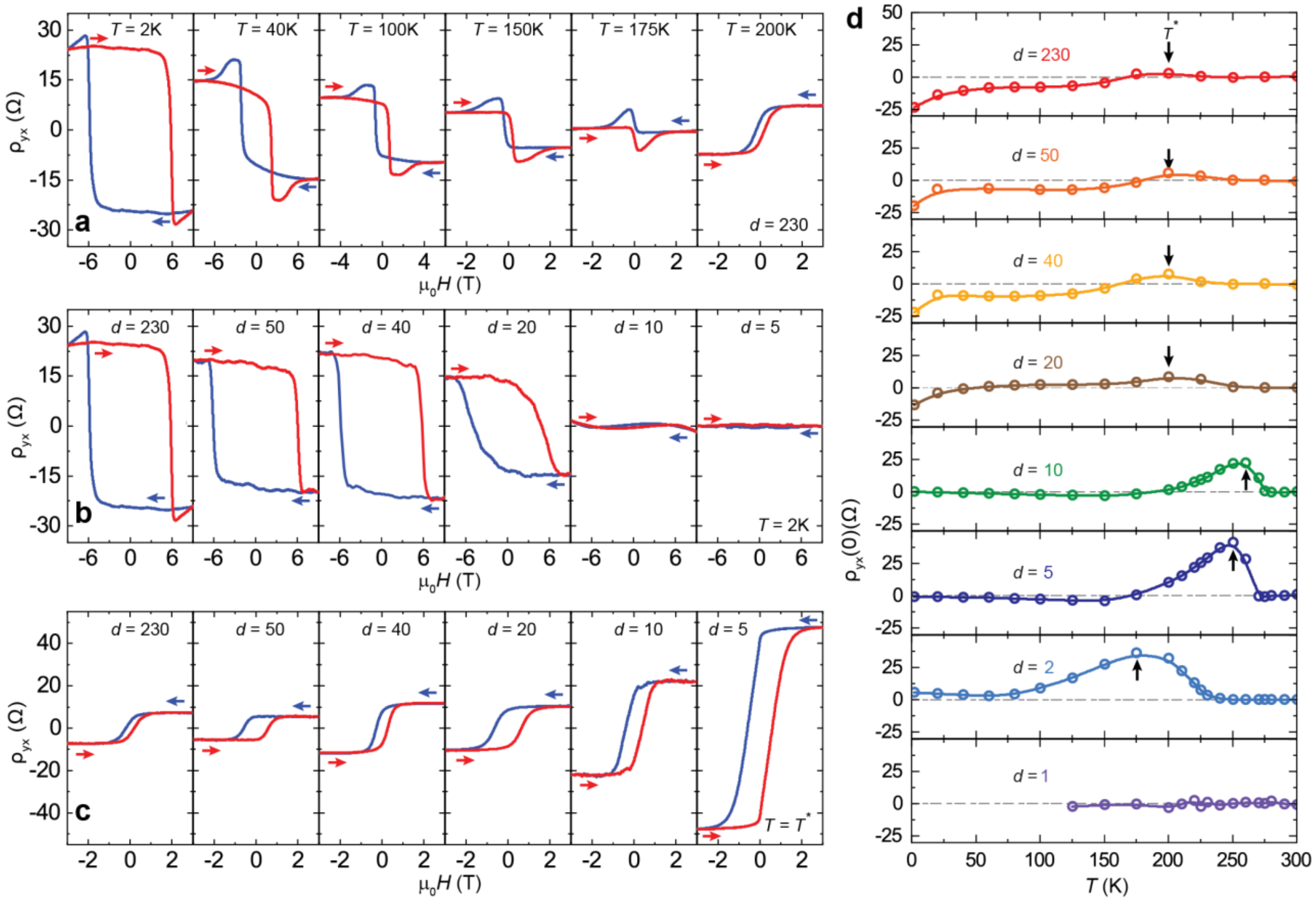

**Fig. 4 | *d*-dependent AH effect in epitaxial MnTe films. a**, $\mu_0 H$-dependent $\rho_{yx}$ measured at $T = 2$ K, 40 K, 100 K, 150 K, 175 K, and 200 K for a MnTe film with $d = 230$. **b**, **c**, $\mu_0 H$-dependent $\rho_{yx}$ for MnTe films with $d = 230$, 50, 40, 20, 10, and 5, measured at $T = 2$ K (**b**) and $T = T^*$ (**c**). $T^*$ denotes the temperature at which $\rho_{yx}(0)$ reaches its maximum. **d**, $T$-dependent $\rho_{yx}(0)$ for MnTe films with $d = 1$, 2, 5, 10, 20, 40, 50, and 230. The arrows indicate the $T^*$ values.

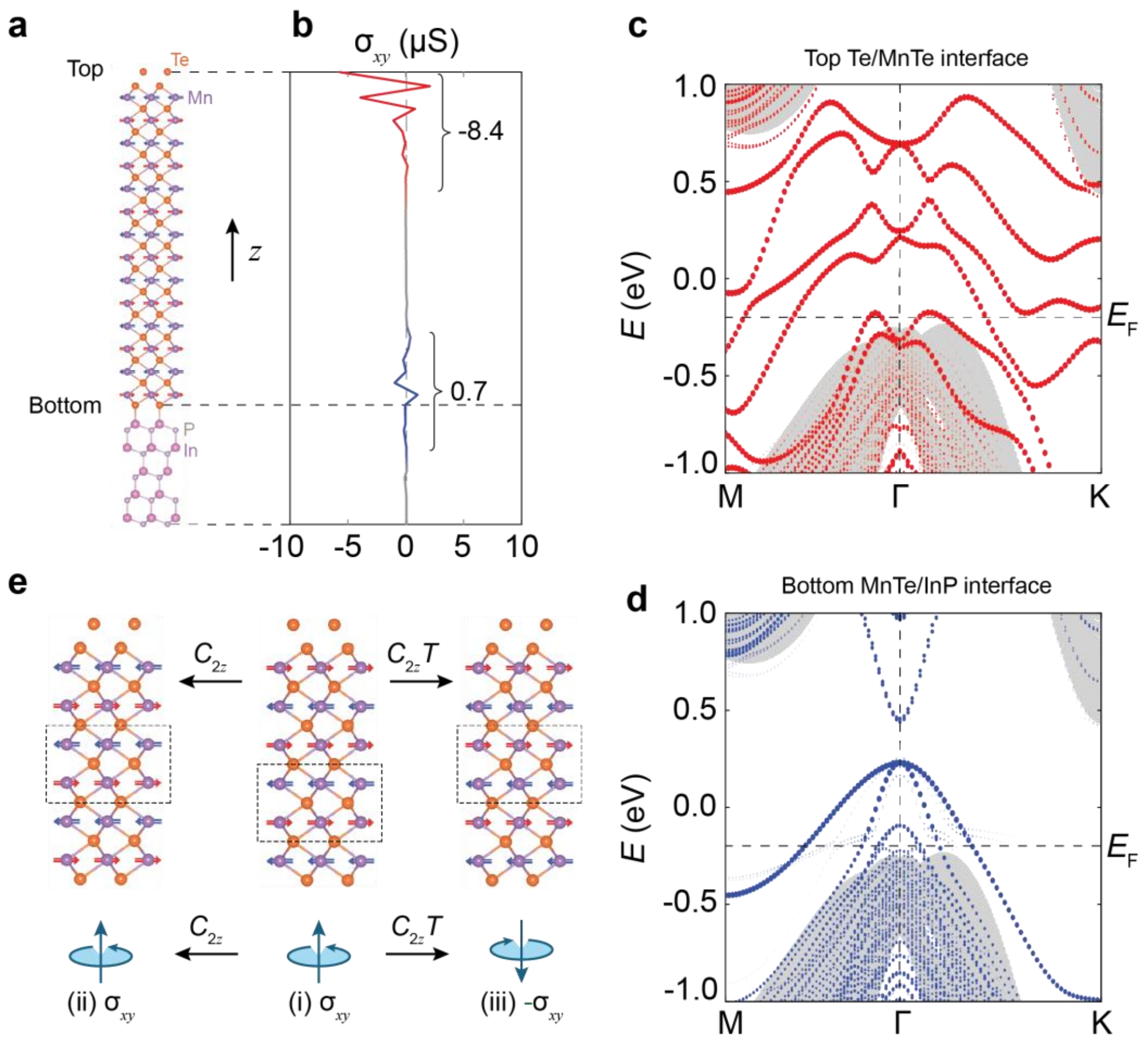

**Fig. 5 | Surface-state-driven AH effect and bulk altermagnetism in MnTe films. a,** Schematic of a MnTe film with $d = 7$ on an InP substrate with a Te capping layer. **b**, Layer-projected $\sigma_{xy}$ along the $z$ direction calculated at $E_F$. The top and bottom regions contribute -8.4 μS and +0.7 μS to $\sigma_{xy}$, respectively. $E_F$ is indicated by the dashed line in (**c**, **d**). **c**, Projected band structure of the top Te/MnTe interface. **d**, Projected band structure of the bottom MnTe/InP interface. The gray shaded regions in (**c**, **d**) represent the bulk states. $E_F$ is set at $E \approx$ -0.2 eV, as determined from ARPES measurements. **e**, Schematic of the MnTe film (i) and its symmetry-related counterparts under the $C_{2z}$ (ii) and $C_{2z}T$ (iii) operations. $\sigma_{xy}$, as a pseudovector along the z direction, is invariant under $C_{2z}$ but changes sign under $C_{2z}T$. $C_{2z}$ denotes the two-fold rotation operation in the magnetic space group, and $T$ is the time-reversal operation. The bulk altermagnetic order is highlighted in the dashed boxes. Configurations (i) and (ii) share the same altermagnetic order and $\sigma_{xy}$, despite reversed magnetization in the terminated MnTe layer, whereas Configurations (i) and (iii) have the same surface magnetization but have opposite bulk altermagnetic order and $\sigma_{xy}$.